\documentclass{aa}  

\usepackage{graphicx}
\usepackage{txfonts}
\usepackage{hyperref}
\begin{document}

   \title{On the feasibility of constraining the triaxiality of the Galactic dark halo with orbital resonances using nearby stars}


\titlerunning{The triaxiality of the Galactic dark halo}

   \author{Casper Hesp \inst{1,2}
          \and
          Amina Helmi \inst{1}}
          
   \institute{
   Kapteyn Astronomical Institute, University of Groningen,
              Landleven 12, 9747 AD Groningen, The Netherlands 
   \and
	Gravitation Astroparticle Physics Amsterdam (GRAPPA), University of Amsterdam, Science Park 904, 1098 XH, The Netherlands \\     
	\email{hespcasper@gmail.com}\\}

   \date{Received April 8, 2018; accepted ?}

 
  \abstract
{It has recently been proposed that if the Galactic dark matter halo were triaxial it would induce lumpiness in the velocity distribution of halo stars in the Solar Neighbourhood through orbital resonances. These substructures could therefore provide a way of measuring its shape.}
   {We explore the robustness of this proposal by integrating numerically orbits starting from a realistic set of initial conditions in dark halo potentials of different shape.}
   {We have analysed the resulting velocity distributions in Solar neighbourhood-like volumes, and have performed statistical tests for 
the presence of kinematic substructures. Furthermore, we have characterized the particles' orbits using a Fourier analysis.}
   {The local velocity distributions obtained are relatively smooth, statistically consistent with being devoid of substructures even for a dark halo potential with significant but plausible triaxiality. Although resonances are indeed present and associated with specific regions of velocity space, the fraction of stars associated to these is relatively minor. The most significant imprint of the triaxiality of the dark halo is in fact, a variation in the shape of the velocity ellipsoid with spatial location.}
   {}


   \maketitle
%

\section{Introduction}

Pure N-body simulations of the formation of dark matter halos in the
concordance cosmological Lambda Cold Dark Matter ($\Lambda$CDM) model
have consistently predicted Milky Way-size halos to be triaxial
\citep[e.g.,][]{allgood06}.  Observationally the determination of the
shape of halos surrounding galaxies is very challenging and
measurements are therefore sparse. Even for the Milky Way, obtaining 
constraints on the shape and orientation of its dark matter halo 
has proven to be tricky. The largest obstacle so far has
been the lack of large samples of stars in the stellar halo with
accurate positions and velocities. This is changing 
dramatically thanks to the Gaia mission
\citep{perryman01,prusti2016} and its (upcoming) data releases \citep[see e.g.][DR1]{brown2016}.

Nonetheless, different attempts have been made to measure the shape of
the dark matter halo of the Milky Way, mostly using the Sagittarius
stellar streams. Early analyses of the spatial location of stars in the streams
favoured slightly oblate shapes
\citep[e.g.][]{Ibata2001,MD2004,kvj2005}, while the velocities of
stars in the leading arm of the stream required a prolate shape
elongated in the direction perpendicular to the Galactic plane
\citep{helmi04}. This tension was resolved by \citet[][]{law10} who were the 
first to explore the possibility that the halo might be triaxial and
actually found when fitting that it appears to be nearly oblate but
with the minor axis pointing towards the Sun. This contrived model
however, can be modified to allow for a flattening that varies with
radius, such that in the central regions it is oblate towards 
the disk, as shown by \citet{vera13}, and as one might expect from physical arguments. Furthermore, these authors have also shown that the
Magellanic Clouds could have a significant dynamical effect on the
stream, such that the underlying Galactic dark matter halo would
actually be triaxial at large radii, with the intermediate axis
pointing in the direction of the Clouds, and with the major axis at
large distances perpendicular to the disk. These results provide
tentative support for the cosmological prediction that halos are triaxial. 

In search of more ways to constrain the Galactic dark halo's
triaxiality, \citet{nino12} have proposed a method based on orbital
resonances. These can induce substructures in the velocity
distribution of e.g. nearby halo stars, which if observable could
potentially be the `smoking gun' of halo triaxiality. Simulations by
these authors showed that such effects might indeed be present.
However, for the sake of computational efficiency the initial
conditions chosen to represent the orbital structure of the Galactic
stellar halo were not sufficiently general, and seemed to overselect
stars populating resonances. In a second more recent study,
\citet{nino15} made an attempt to reduce the selection bias but still
used initial conditions uniform in position, while observationally the
density of the halo is known to fall-off steeply with
radius \citep{helmi08}. Therefore, the question remains as to whether orbital
resonances leave an important imprint on the local halo velocity
distribution and hence provide an observable constraint on the triaxiality of
the dark halo \citep[see also][]{Valluri2012}. The purpose of the current study is to address this
point with more realistic orbital integrations.

The general outline of this paper is as follows. In Section
\ref{section:methodinit} we describe how we generate initial
phase-space coordinates for a self-consistent spherical distribution
of halo stars. These are then integrated in triaxial, oblate, and
prolate dark halo NFW-potentials \citep{NFW96} whose characteristics
are defined in the same Section. We also perform additional orbit integrations including a disk component. After integration, we select
stars from three different Solar neighbourhood-like regions, located
on the major, intermediate, and minor axes of the dark halo. For these
stars, we quantify the significance of substructures in velocity space
in Section \ref{section:velanalysis}. In Section
\ref{section:resanalysis} we identify the different orbital
resonances, and link back this information to the velocity
distributions. In Section \ref{section:discuss} we discuss our results
and contrast them with those of \citet{nino12} and \citet{nino15},
while in Section \ref{section:conclusions} we summarize our
conclusions.\section{Set-up and Initial Conditions}
\label{section:methodinit}

\subsection{Initial Conditions}
\label{section:ic}

We aim to generate initial conditions that are a more realistic
representation of the six-dimensional phase-space structure of the
stellar halo of the Milky Way. As a starting point and for simplicity,
we assume the distribution function to be a function of energy $E$ and
angular momentum $L$: $f(E,L) = g(E) h(\eta)$, where $\eta = L/L_c(E)$
is the circularity parameter. We also assume that the stellar halo is
a power-law tracer population embedded in a spherical potential,
representing the dark matter halo of the Galaxy. 

For the potential we choose the NFW form:
\begin{equation}\label{eq:NFW}\begin{aligned}
\Phi_{NFW}(r) &= - \frac{GM_{200}}{r_sf(C_{200})}\frac{\ln{(1+r/r_s)}}{r/r_s},
\end{aligned}\end{equation}
where $f(u) = \ln{(1+u)}-u/(1+u)$, the scale radius $r_s = 18$~kpc,
$M_{200}$ is the virial mass, which we set to $\sim8.35 \cdot 10^{11}
M_{\bigodot}$ \citep[close to the value for the Milky Way,
e.g.,][]{battaglia06, busha11, caut14}.  The concentration parameter
$C_{200} = r_{200}/r_s \sim 10.82$. 

To assign orbital energies to the stars we set their apocenter
distances, $r_a$, such that $ p(r_a)dV \propto \rho(r_a)dV \propto
r_{a}^{-3.5}dV$, i.e. following the number density
distribution of halo stars $\rho
\propto r^{-3.5} $ \citep{kinman94}. After setting the inner radius to $r_c = 0.3$~kpc to avoid
divergence of the integral, we integrate this equation to find the
normalization factor. The resulting probability distribution for $r_a$
is:
\begin{equation}\begin{aligned}
    p(r_a)dr_a &=
    \frac{1}{2}\sqrt{\frac{r_c}{r_a^3}}dr_a. \end{aligned}\end{equation}
From this distribution we draw the apocenter radii with the extra
condition that $3$ kpc $ < r_a < 100$ kpc to avoid orbits
that are too close or too far to be relevant for our analysis which
concerns the Solar Neighborhood at $\sim 8$ kpc from the Galactic
center. To ensure spherical symmetry the angles are distributed isotropically, i.e.
uniformly in $\cos{\theta}$ and $\phi$.

We set the particles initially at their orbital apocentres. To determine their (tangential) velocity we 
first derive the angular momentum of a circular orbit
$L_c$ at a given $r_a$ using the NFW potential of Eq.~\eqref{eq:NFW}. Then the
circularity parameter $\eta \equiv L /L_c(r_a)$ is randomly extracted from the
distribution derived from cosmological simulations by
\citet{wetzel11}. Finally the tangential velocity $V_t = L/r_a = \eta L_c(r_a)/r_a$. 

We use this method to produce six-dimensional phase-space coordinates
for $2 \cdot 10^6$ test particles. To check the correctness of this
approach we first integrate these initial conditions in the spherical
NFW potential assumed in the set-up.  After phase-mixing we find that
the number density distribution follows indeed the power law $\rho
\propto r^{-3.5}$. Inside spherical volumes of 2 kpc radius located at
a distance of 8 kpc from the center of the halo (that could represent
our Solar Neighborhood), the kinematics are well fit by Gaussians with
$\bar{V_r} = \bar{V_\theta} = \bar{V_\phi} \sim 0$~km/s, $\sigma_{r}
\sim 70$~km/s, and $\sigma_{\theta, \phi} \sim 80$~km/s. Note that
these velocity dispersions are smaller than the dispersions for halo
stars in the Solar Neighborhood \citep[see observations summarized
by][]{kepley07}. This is not unexpected since the circular velocity at
8~kpc in our model is $\sim 106$~ km/s as we have not included the disk
component of the Milky Way which is an important mass contributor near
the Sun.

\subsection{Non-spherical NFW Potential}\label{section:methodNFW}

\citet{vog07} have proposed a triaxial generalisation of the spherical NFW potential.  In Equation \eqref{eq:NFW} the radius $r$ is replaced by a generalized radius $\tilde{r}$:
\begin{equation}\label{eq:rtilde}\begin{aligned}
\tilde{r} = \frac{(r_\alpha+r)r_E}{(r_\alpha + r_E)}
\end{aligned}\end{equation}
where $r_E$ is an ellipsoidal radius:
\begin{equation}\label{r_E}\begin{aligned}r_E = \sqrt{\frac{x^2}{a^2}+\frac{y^2}{b^2}+\frac{z^2}{c^2}}.\end{aligned}\end{equation}
The condition for normalization is
that $a^2+b^2+c^2=3$. In Eq.~\eqref{eq:rtilde} $r_\alpha$ is the transition scale which we take to be $1.2 r_s$. 
Note that for $r \ll r_\alpha$ then $\tilde{r} \sim r_E$, while in the outer regions, $\tilde{r} \sim r $. 

For the sake of generality we consider three different shapes: a
prolate halo, with the major axis pointing in the $z-$direction \citep[$a/c \approx b/c \approx 0.6$, i.e. more flattened than proposed by][]{helmi04}, an oblate halo with the minor axis in the $z-$direction 
\citep[$a/b\approx0.99$, $c/a\approx0.72$;][]{law10}, and a triaxial
halo \citep[$a/b\approx0.83$, $a/c\approx0.67$; used
by][]{nino12}. The axial ratios chosen for the triaxial halo are
more extreme than the average values found in cosmological N-body simulations
of Milky Way-like dark matter halos \citep{vera11}, but still
realistic. Note that these are all axial ratios for the potential and
not for the density distribution which is flatter.
	
\subsection{Including a Miyamoto-Nagai Potential}\label{section:methodDisk}

The inclusion of a disk component would increase the realism of these
simulations. However, it would also increase the axial symmetry
of the system which could potentially reduce resonant effects caused by the
triaxiality of the halo. Therefore, our main analysis does not involve
a disk component which allows us to obtain an upper bound on its
effects. On the other hand, the low velocity dispersions obtained by
considering only the dark halo mass contribution could possibly reduce
the number of particles that populate orbital resonances if these were
more likely at larger velocities. Therefore, we perform 
additional simulations which include a disk component to explore the differences.

We generate initial conditions as described in Sec.~\ref{section:ic},
but now for an NFW halo with $M_{200} = 2.5 \cdot 10^{12} M_\odot$ ($C_{200} = 15.6$ and $r_s = 18$~kpc).
These initial conditions are then integrated using either a spherical
or triaxial NFW and a disk component.  The disk is placed on the $x$-$y$
plane and follows the Miyamoto-Nagai potential \citep{miyamoto75}:
 \begin{equation}\label{eq:miyamoto}\begin{aligned}
\Phi_{disk}(x,y,z) = \frac{-GM_{disk}}{\sqrt{x^2+y^2+(A+\sqrt{z^2+B^2})^2}},
\end{aligned}\end{equation}
where the scale radius $A = 5.32$ kpc and the vertical length scale $B
= 0.25$ kpc as derived by \citet{allen91} for the Milky Way. We set
$M_{disk} = 5\cdot 10^{10} M_\odot$, which yields a circular velocity on the disk plane at 8 kpc of $\sim 242$~ km/s. Integration of the initial conditions in
a potential containing a spherical NFW and a disk component as
described in Sec.~\ref{section:ic} leads to velocity dispersions in
the Solar Neighborhood-like volumes whose amplitude is in better agreement with
observations of local halo stars \cite[for volumes located in the
plane of the disk: $\sigma_{r} \sim$ 120~km/s, $\sigma_{\theta, \phi}
\sim$~143 km/s; see e.g.,][]{kepley07}. 

\section{Stellar Kinematics: Results and Analyses} \label{section:velanalysis}
\subsection{Overall Local Velocity Distributions}\label{section:resultvel}

\begin{figure}[!h]
\centering
\includegraphics[height=14cm]{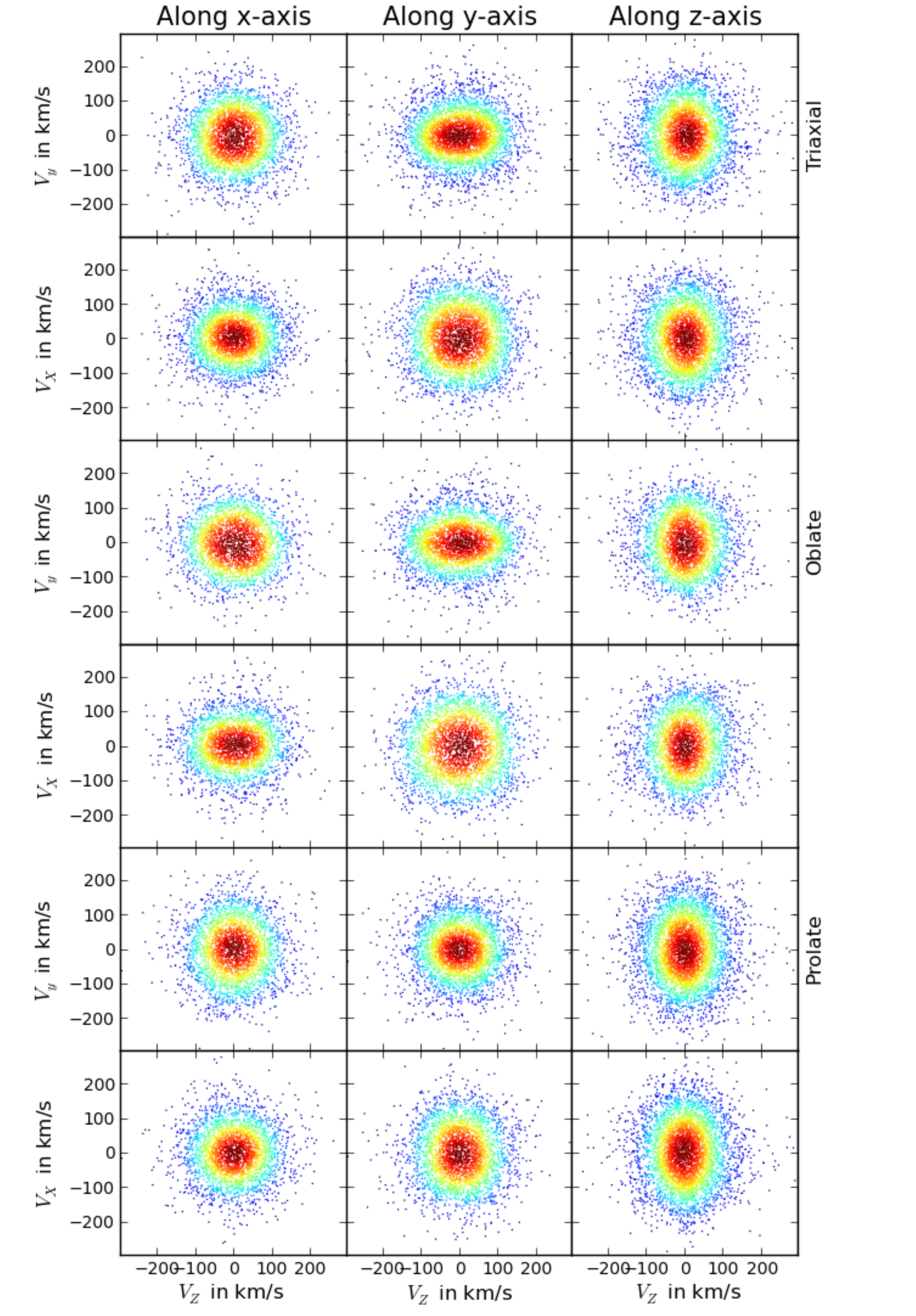}
\caption{Velocity distributions inside three different locations of the Solar Neighborhood volumes: along the $x$-, $y$- and $z$-axis, obtained after integration for $8 $ Gyr for the triaxial (top two rows), oblate (middle two rows), and prolate (bottom two rows) dark matter halo.  In all cases is the major axis is aligned with the $z$-direction.}
\label{fig:onaxisvelxyz}
\end{figure}

We integrate the orbits of $2 \cdot 10^6$ particles for 8 Gyr in the
dark halo NFW potentials for three different shapes, i.e. triaxial,
oblate, and prolate. In Figure \ref{fig:onaxisvelxyz}, we show the
resulting velocity distributions within the Solar Neighborhood volumes
(SN) located on the three principal axes. As expected, the
distributions are centered around $\sim 0 $ km/s, and appear to be
smooth. The eigenvectors of the velocity ellipsoids are aligned with
the principal axes of the dark halo. The largest dispersions are found
in the plane perpendicular to the major axes (triaxial, oblate,
prolate: $\sigma_{\perp major} \sim 85$~km/s), and while the smallest
are along the direction of the major axes (triaxial:
$\sigma_{\parallel major} \sim 70$~km/s, oblate: $\sigma_{\parallel
  major} \sim 75 $~km/s, prolate: $\sigma_{\parallel major} \sim
66$~km/s). In the slightly off-major or -minor axis volumes ($\sim$14
degrees), the velocity ellipsoids are tilted $\sim$14 degrees.  For
the simulations involving a disk the velocity ellipsoids are similarly
smooth. In the case of the spherical NFW with a disk, $\sigma_{r} =$
121~km/s, $\sigma_{\theta, \phi} \sim$~143 km/s on the plane. For the
triaxial NFW with a disk component, the ellipsoid on the $x$-axis is
$(\sigma_x,\sigma_y,\sigma_z) = (129, 131, 134)$~km/s, while on the
y-axis it is $(139,124,132)$~km/s. Such variations with location of
the properties of the velocity ellipsoid could perhaps be used with
the aid of e.g.  the Jeans equations \citep{b&t08} to constrain the
shape of the halo \citep[see also][]{Smith2009}.

As can be seen from Figure \ref{fig:onaxisvelxyz} no clear
substructures are apparent in the velocity distributions (and this is the case whether a disk component is included or not).

\subsection{Kullback-Leibler Divergence} \label{section:methodKLD}

To quantify further  whether there are any significant substructures in the velocity distributions, 
we apply the Kullback-Leibler Divergence (hereafter:
KLD). The KLD is a statistic that measures the 
relative entropy or information content of two distributions, and can be used to determine the relative degree of
clumpiness. We use the KLD to compare the velocity distribution $P(i,j,k)$ found for the particles in our simulations sampled on a three-dimensional
cartesian grid in velocity space $(i,j,k)$ to a smooth 
velocity distribution $Q(i,j,k)$ computed on the same grid. For discrete probability distributions in three
dimensions we define the KLD statistic $\alpha$ as:
\begin{equation}\begin{aligned}
\alpha = \sum_{i,j,k} P(i,j,k)\ln{\left(\frac{P(i,j,k)}{Q(i,j,k)}\right)}.
\label{KLD}\end{aligned}\end{equation}
The larger the value of $\alpha$, the more clumpy the distribution. 
We use three different bin sizes, $25 $ km/s, $50 $ km/s, and $100 $
km/s, loosely tuned to the size of the orbital structures (highlighted in
the next Section). The total number of particles in the volume is typically $N_{tot} \sim 3000$.

The smooth velocity distribution used for comparison, $Q(i,jk)$, is obtained by randomly shuffling the velocity distribution found in the simulations. We couple each value of $v_x$ with a randomly selected $v_y$ and $v_z$ (without replacement). In this way, we generate 5,000 shuffled data sets and average the binned values to obtain $Q(i,j,k)$, making sure that any correlations and clumpiness are broken \citep[see e.g.][]{Sanderson2015}

To establish the statistical significance of a given value $\alpha$ found in our simulations,
we determine how likely it is to obtain it in the shuffled data sets. We construct a distribution of $\alpha_{shuffle}$ by computing, for each of the 5,000 shuffled data sets, the $\alpha$ value using Equation
\eqref{KLD}. We may now calculate the probability $p_{shuffle} = N(\alpha_{shuffle} \ge
\alpha_{sim})/N_{tot}$ of finding a value of $\alpha_{shuffle}$ as
large as or larger than the measured $\alpha_{sim}$. To obtain more robust results, we have analysed the orbital integrations at 7 different outputs, starting from 8 Gyr up to 20 Gyr in total. 

The vast majority of the volumes have median $p_{shuffle} > 0.1$ for all potentials, whether axisymmetric or triaxial, demonstrating that on average, there are no significant differences in the amount of kinematic substructure present. Every now and then, a volume might depict a low value of $p_{shuffle} \sim 0.001$. This happens for the oblate and for the triaxial halos with similar frequency, indicating that, although the kinematic distribution might show some degree of lumpiness (on scales of 50 km/s), this is not necessarily a signature of triaxiality.

\section{Orbital Analyses}\label{section:resanalysis}
\subsection{Characterization of Orbital Resonances} \label{section:methodorb}

Orbits that are quasi-periodic can be expressed as:
\begin{equation}\label{eq:regulardef}\begin{aligned}
\vec{q}(t) = \sum_{k=1}^{\infty}\vec{a_k}e^{iF_k t}
\end{aligned}\end{equation}
where $\vec{q}(t)$ is the vector of the spatial coordinates,
$\vec{a_k}$ the vector of the amplitudes, and the frequencies $F_k$
are linear combinations of $N_b$ base frequencies. Therefore the
Fourier transform of a quasi-periodic orbit will consist of a sum of
peaks with amplitudes $\vec{a_k}$. In general, one is interested in
determining the number and the numerical values of the base
frequencies $\vec{F_{b}}$. Orbits in a triaxial potential probe three-dimensional
regions of space and therefore most will have three base-frequencies,
$N_b = 3$, unless the frequencies are commensurate, i.e. they are on a
resonance. On the other hand, there may be orbits that are irregular
and so cannot be described by Equation \eqref{eq:regulardef}. In
practise this might be taken to mean that the frequencies found by assuming
Eq. \eqref{eq:regulardef} will be (combinations of) more than 3
frequencies, and hence these cannot be properly called base
frequencies. 

We apply the spectral orbit classifier of \citet{carp98} to identify
the base frequencies of the orbits in our simulations.  After
obtaining the time-series of the orbits through numerical integration, we identify
the dominant frequencies $F_{d}$ by taking the frequencies with the
highest amplitudes $a_k$. In all cases, we use the time-series in
cartesian coordinates, i.e. $x$, $y$, and $z$. Accordingly, we find
three groups of dominant frequencies $\vec{F_{x,d}}$, $\vec{F_{y,d}}$,
and $\vec{F_{z,d}}$ for each orbit where we consider a maximum of five
per group. Through a process of subtracting the dominant frequencies
and multiples thereof from the power spectra and reanalyzing the
spectra, the orbit classifier of \citet{carp98} determines the number
$N_b$ of fundamental frequencies $F_b$ that describe the motion of
each particle.  

The so-called `frequency maps' \citep{b&t08} are obtained by plotting
$F_{2,b}/F_{3,b}$ versus $F_{1,b}/F_{3,b}$ help to visualize the
structure in frequency space \citep[e.g.][]{Valluri2012,Valluri2013}.  For regular orbits that are on a
resonance we expect the base frequencies to be coupled in two or three
dimensions. We define an orbital resonance when the ratios of the
base frequencies in two or three dimensions 
is a rational number. This can be expressed by:
\begin{equation}\label{resequation}\begin{aligned}
l_1F_{1,b} + l_2F_{2,b} + l_3F_{3,b} = 0,
\end{aligned}\end{equation}
where $l_1$, $l_2$, and $l_3$ are integers and $F_{1,b}$, $F_{2,b}$,
and $F_{3,b}$ are the base frequencies. Particles on orbital
resonances therefore populate straight lines in the frequency
map.


\begin{figure*}
\centering
\includegraphics[width=17cm]{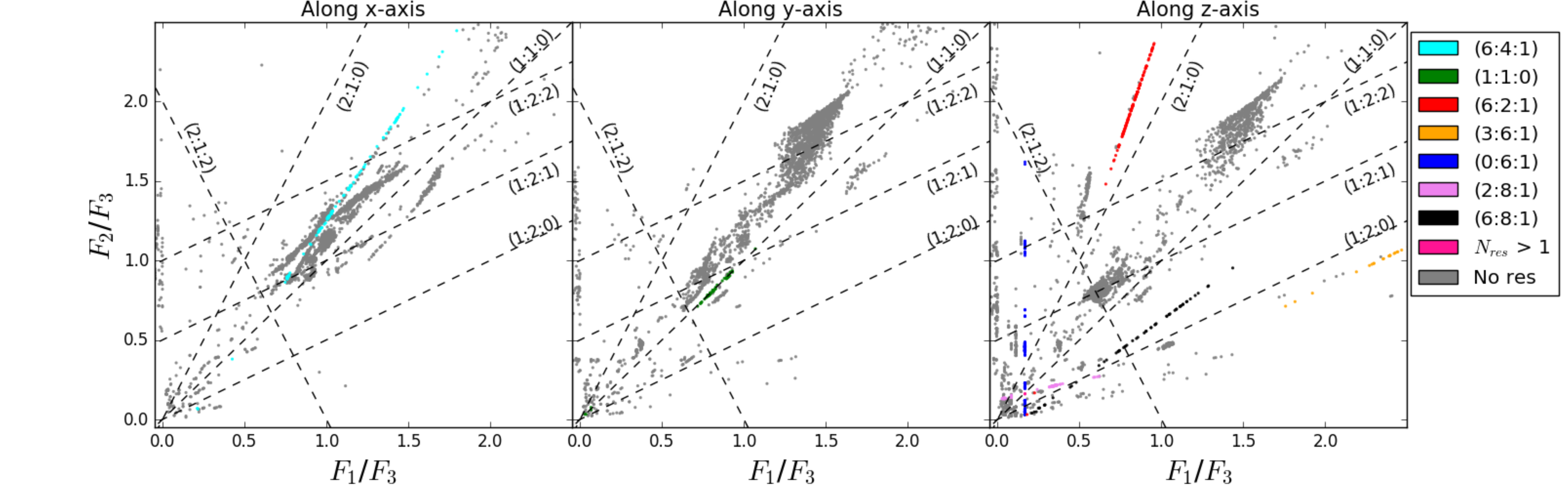}
\caption{Frequency map for $\sim 3000$ particles located inside ``Solar neighbourhood'' spheres 
of radius $2 $ kpc located on the $x$-, $y$-, and $z$-axes, obtained by 
orbital integration within the triaxial NFW. The orbital resonances can be distinguished by lines that follow the equation $l_1F_{1,b} + l_2F_{2,b} + l_3F_{3,b} = 0$ with all $l$ integer numbers and $\vec{F_b}$ the three base frequencies. The conditions are that particles are within a range of +/- 0.01 of a specific resonance line, at least ten particles populate the resonance, the integers $l$ of the resonance are all smaller than 10, and there is a distinct region around the line that is not populated. This last condition is partly violated in the case of resonance [6:4:1] on the $x$-axis, but we decided to include it for exploratory purposes.}
\label{fig:res}
\end{figure*}

\begin{figure*}
\centering
\includegraphics[width=16cm]{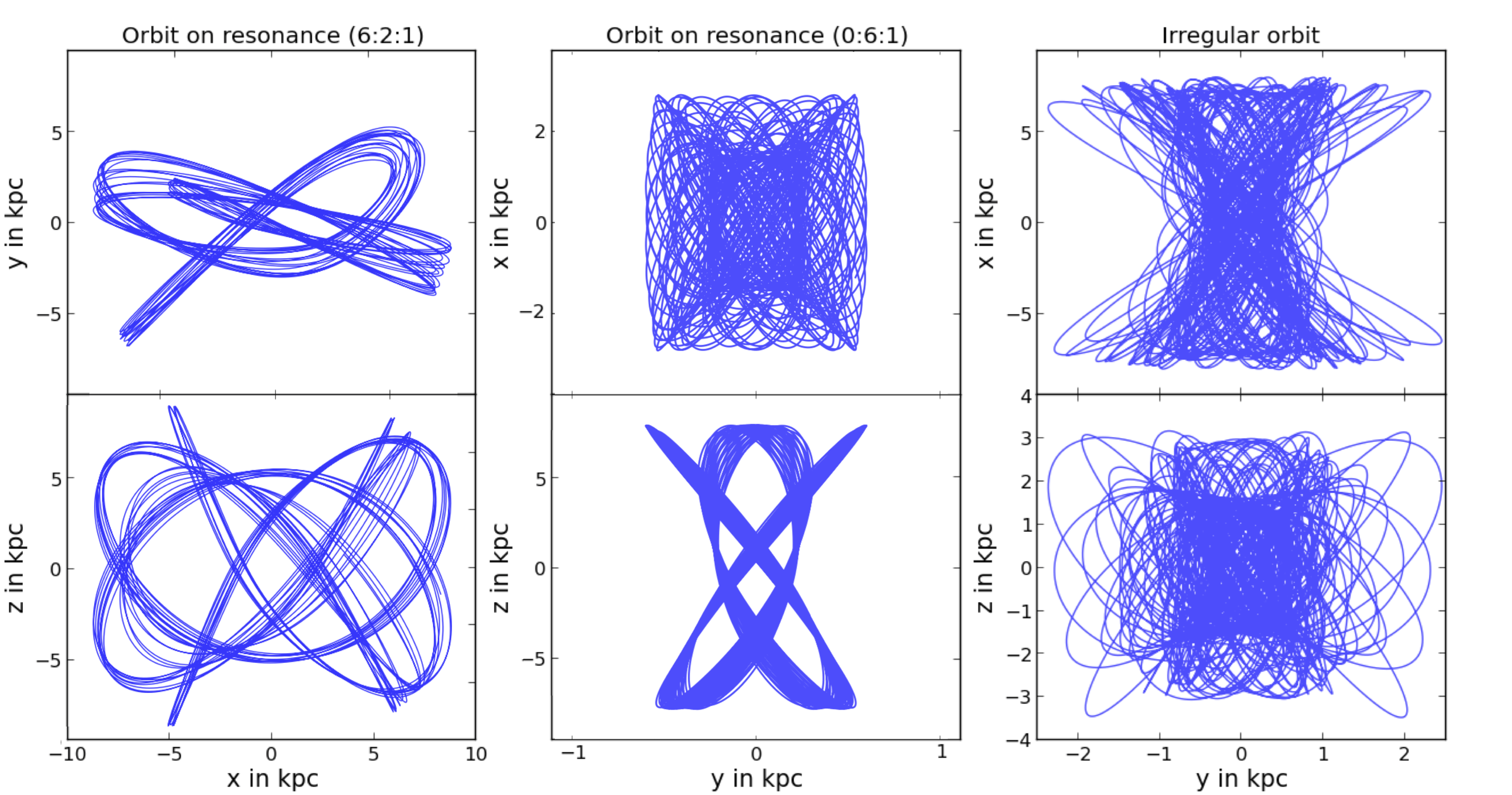}
\caption{Examples of three orbits integrated in the triaxial NFW-potential for 15 Gyr. The left and middle panels correspond to orbits on the resonances [$l_1$:$l_2$:$l_3$] = [6:2:1] and [0:6:1] respectively, while the panel on the right shows an irregular orbit with $N_b >$ 3.}
\label{fig:orbitres}
\end{figure*}

\subsection{Resonance Populations Identified with Fourier Analysis}\label{section:resultfreq}

We analyze the orbits of the particles in the volumes shown in
Figure \ref{fig:onaxisvelxyz} and integrated for $\sim$65 Gyr, using the orbit classifier of
\citet{carp98}. In Figure \ref{fig:res} we plot the resulting
frequency map $F_{1,b}/F_{3,b}$ versus $F_{2,b}/F_{3,b}$ for the
triaxial potential. We use Equation \eqref{resequation} to identify
the particles that populate resonances and the different types of
resonances. This is done under a number of conditions, namely $i$)
that particles are within a range of +/- 0.01 of a specific resonance
line, $ii$) at least ten particles populate the resonance, $iii$) the
integers $l$ of the resonance are all smaller than 10, and $iv$) there
is a distinct region around the line that is not populated. These
resonances are indicated with different colours in Fig.~\ref{fig:res}.
It is evident from this figure that the orbital structure is different
from volume to volume, and that the prominence of resonant orbits also
varies with location. 

We also use the orbit classifier of \citet{carp98} to determine the
fractions of orbits of different types (regular, resonant, irregular)
in the various volumes. Figure~\ref{fig:orbitres} shows some examples of the
types of orbits found in the triaxial potential.  Table
\ref{tab:fracreg} summarizes the overall findings regarding the
fractional distribution of the irregular orbits, non-resonant regular
orbits, the resonant orbits (coupled in two or three dimensions), and
the thin orbits (i.e., orbits for which the orbit classifier only
found two base frequencies). On the major axis ($z$) in the triaxial
potential $\sim$18\% of the particles populate resonant orbits. More
than half of these particles are on orbits that are coupled in three
dimensions, they populate the resonances [$l_1$:$l_2$:$l_3$] =
[6:2:1], [3:6:1], [2:8:1], and [6:8:1]. The remainder resonant orbits
are coupled in only two dimensions. For the volumes on the
intermediate and minor axes we see a much smaller fraction of resonant
orbits ($\sim$2--4\%).  Note as well that in the volume on the major
axis there is a much larger fraction of irregular orbits ($\sim$ 25\%)
compared to the intermediate and minor axes ($\sim$ 5\%).

Similar analyses
for two volumes $\sim$14 degrees off-axis from the major and minor
axes showed orbital populations comparable to those found on the
principal axes closest to these volumes.

The oblate and prolate potentials showed only resonances coupled in
two dimensions. Again the volumes on the major axes show larger
fractions of resonant orbits than those on the minor axes. The
fraction of irregular orbits in the various volumes are generally much
smaller in these potentials (oblate: $<$ 12\%; prolate: $<$ 4\%).

In summary, we note that in each volume the fractions of resonant
orbits and irregular orbits are very small compared to the fraction of
non-resonant regular orbits. Only on the major axis of the triaxial
potential the fractions of both resonant and irregular orbits are higher (each contribute close to 20\%). If present, the resonances are always of high order.

\begin{table}
\caption{The fractional distribution of non-resonant orbits (regular and irregular), resonant orbits (coupled in two or three dimensions) and thin orbits on the $x$-, $y$-, and $z$-axes in the triaxial NFW. The $x$-, $y$- and $z$-axes are aligned with the minor, intermediate, and major axes of the potential.}
\label{tab:fracreg}
\centering
\begin{tabular}{|ll|l|l|l|}
\hline
Triaxial     &                 & x    & y    & z    \\ \hline
Non-resonant & Regular (\%)    & 88.6 & 91.8 & 57.3 \\ \cline{3-5} 
             & Irregular (\%)  & 6.6  & 4.1  & 23.1 \\ \hline
Resonant     & 2D-coupled (\%) & 0.6  & 1.1  & 7.9  \\ \cline{3-5} 
             & 3D-coupled (\%) & 3.5  & 0.9  & 10.1 \\ \hline
Thin(\%)     &                 & 0.8  & 2.0  & 1.6  \\ \hline
\end{tabular}
\end{table}

\subsection{Resonance Populations in Velocity Space}\label{section:velres}

\begin{figure*}
\centering
\includegraphics[width=16.5cm]{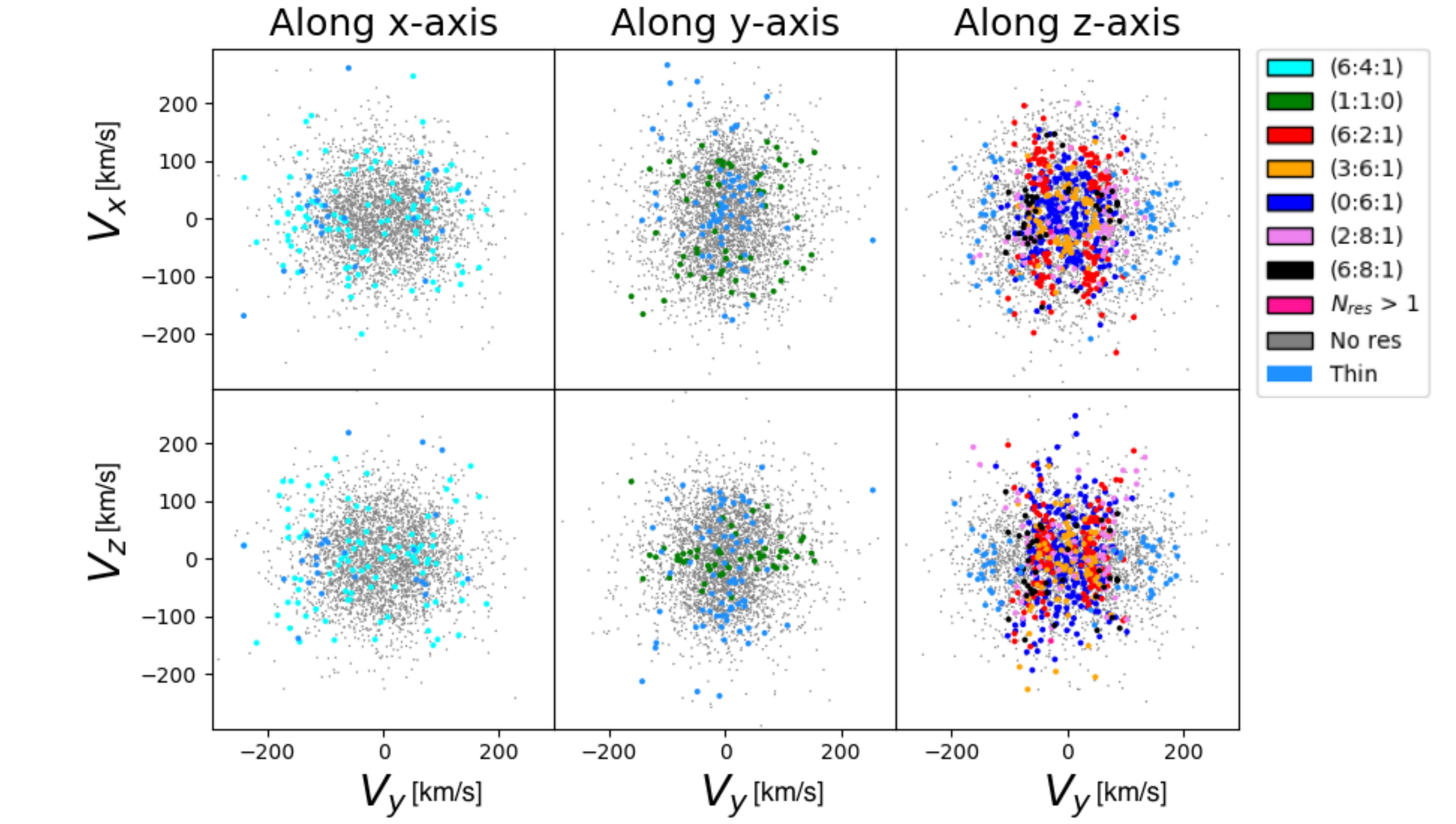}\par
\caption{Velocity distributions inside the volumes along the $x$, $y$, and $z$-axes with the orbital resonances 
represented by different colors as in Fig.~\ref{fig:res}.}
\label{fig:velrestriax}
\end{figure*}

Now we link the information of the resonance groups from Section
\ref{section:resultfreq} back to the velocity distributions of the
particles for the selected Solar neighbourhood volumes. The resonant
populations identified in frequency space are shown in Figure \ref{fig:velrestriax} with the same
colour coding in velocity space for
the triaxial potential. We see that for the volumes located on and
near the major axis $z$ different resonances appear to populate different
cylindrical shells in velocity space, all varying in their
orientation. For example, the resonance [$l_1$:$l_2$:$l_3$] = [3:6:1]
(in orange) forms a small well-defined ``ring" at $|V_{x,y}| \sim 50$
km/s in the top right panel of Figure \ref{fig:velrestriax}, while the
resonance [6:2:1] (in red) forms two ``walls" which have the range of +/-
150 km/s in $V_x$ and $V_z$ centered on $V_y$ = 50 km/s and $-50 $
km/s. The particles on thin orbits populate mainly the regions of
$|V_y| > 100 $ km/s. On the $x$- and $y$-axes much smaller fractions
of the particles populate resonances and the patterns are not as
pronounced as on the $z$-axis. Nevertheless, some features can be
distinguished. For example, the resonance [6:8:1] forms a narrow line
at $V_x = V_z = 0 $ km/s on the $y$-axis and particles from the
resonance [6:4:1] mostly populate a bar-shaped region of thickness 50
km/s centered on $V_z$ = 0 km/s. 

Similar behaviour is found for two volumes $~$14 degrees off-axis for
the major and minor axes, as well as for the orbital integrations
including a disk component, where regions in velocity space can be 
related to orbital resonances. \par

\section{Discussion}\label{section:discuss}

The analysis presented above does not support the main claim of
\citet{nino12} that orbital resonances should be readily apparent in the
velocities of nearby halo stars and therefore could be used for
constraining the shape of the Milky Way's dark matter halo. Although
certain regions in the velocity distribution can indeed be related to
resonances, we find in our model that the prominence of these features and the associated number
of stars is very small. The reason for this disagreement can be
explained by closer inspection of the methods used for setting up the orbital
initial conditions.

\begin{figure}
\centering
\includegraphics[width = 8cm]{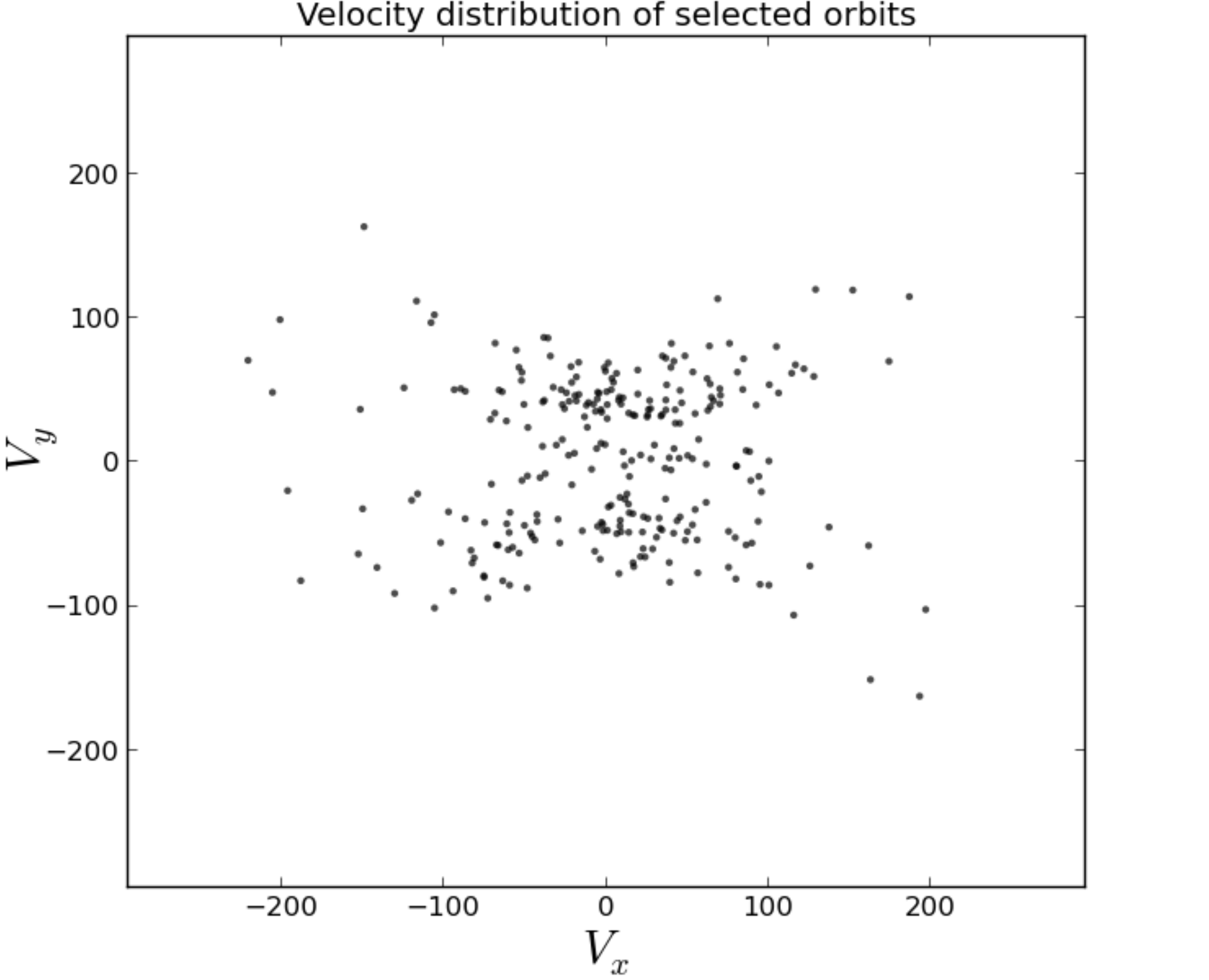}
\caption{Velocity distribution of the particles located in a Solar Neighborhood volume of radius $1.5 $ kpc at 
8.5 kpc from the center on the $x$-axis after selecting only the particles that populate the resonances ($l_1$:$l_2$:$l_3$) = (6:2:1), (3:6:1), and (2:8:1). Although substructure is clearly apparent, most particles in this volume are on regular non-resonant orbits which yields velocity distributions as those shown in Fig.~\ref{fig:velrestriax}.}
\label{fig:nino}
\end{figure}

In their set-up, \citet{nino12} integrate the positions and velocities
of stars initially located in a 1~kpc Solar-neighbourhood-like sphere, and
focus on the kinematical structure {\it in the same
  volume} after 12 Gyr of evolution. This effectively pre-selects
particles on resonant orbits, since those not associated to any
resonance are less likely to return to the same volume on a
relatively ``short'' timescale. This crucial difference explains why
we find that the vast majority of orbits in such volumes are
regular but non-resonant, and why we find smoother 
velocity distributions. To confirm this explanation we show in
Figure \ref{fig:nino} how the velocity distribution in our simulations
would appear if we had pre-selected the particles on resonances coupled in three dimensions (i.e., most likely to return to the
same volume). The resulting pattern is similar to that shown by
\citet{nino12} in their Fig.~3. 

More recently \citet{nino15} attempted to reduce this
selection bias by increasing the volume of the sphere where the
initial conditions were sampled. For example, the positions were
generated uniformly within a 25~kpc radius sphere, and the velocities
were sampled randomly between zero and the escape velocity at each
location. This clearly implies a density distribution that does not
fall off with radius, which is unlike that observed for stellar halos, and which presumably oversamples
the outer regions of the velocity distribution, perhaps leading to an
overpopulation of resonant orbits \citep[see, e.g.,][]{merrit99}. 

If the stellar halo was built via accretion
\citep{hw1999}, the traces of these events are
expected to be visible as substructures in velocity space
\citep{helmi08,Gomez2013}.  Depending on the infall parameters of these
building blocks and on the gravitational potential of our Galaxy,
the debris could be more evident if on resonant orbits \cite[as argued
by][]{nino12}, or very difficult to detect if on chaotic orbits as the stars will drift away more
quickly \citep{price-whelan2015}. First attempts to quantify the
likelihood of this in a cosmological setting have been reported by
\cite{Gomez2013,Maffione2015}.  In our simulations, the fractions of
irregular orbits are only relatively large in the volumes located near
the major axis of the triaxial potential. This could imply a bias in
the accretion events that are amenable to recovery as substructures in
velocity space. This is also
relevant for the detection of streams in external galaxies, where
those that may be more prone to detection may be those that
are on specific resonant orbits. It will be important to
quantify the likelihood of this and establish the resulting biases for
 constraining the
accretion or merger history of a galaxy \citep[but see, e.g.][]{Maffione2018}.

\section{Conclusions}\label{section:conclusions}

We have conducted numerical experiments with test particles in a Milky
Way-like dark matter halo of three different shapes: triaxial, oblate,
and prolate. To this end we generated initial conditions for these test particles
following a smooth nearly self-consistent distribution function. The
goal was to investigate the assertion of \citet{nino12} that orbital
resonances caused by the shape of the halo would leave imprints in
the velocity space of  Solar Neighborhood halo stars.

On the basis of the detailed analysis of the kinematics and orbital properties of particles located inside Solar neighbourhood volumes along the principal axes of the dark halo we have reached the following conclusions:
   \begin{enumerate}
    \item The shape of the velocity ellipsoid in these volumes is the only evident difference between experiments using triaxial, oblate, prolate, and spherical dark matter halos, whether including a disk component or not.    
\item The velocity distributions within the Solar Neighborhood volumes do not show any significant substructures for either of the potentials explored, including the triaxial NFW with a disk component. 
    \item While resonances are indeed present and related to specific regions in velocity space, the fraction of particles associated to
resonant orbits is small even for extreme values of the axis ratios that are still consistent with cosmological dark-matter-only simulations. 
   \end{enumerate}

Our numerical experiments show that a smooth and more realistic initial distribution function does not favor any particular orbital resonance that can clearly differentiate among plausible shapes for the Galactic dark matter halo.  We conclude
that the most promising way to pin down its shape using the kinematics of halo stars would
be to inspect the changes in the velocity ellipsoids across
the Galaxy. This ought to be feasible with the six-dimensional
phase-space information of stars 
that will soon be provided by the Gaia mission  \citep{prusti2016,brown2016}.

\begin{acknowledgements}
We thank Robyn Sanderson and Maarten Breddels for providing various source codes. CH acknowledges support from the Amsterdam Science Talent Scholarship of the University of Amsterdam. AH gratefully acknowledges financial support from a VICI grant from the Netherlands Organisation for Scientific Research, NWO.
\end{acknowledgements}


\end{document}